# Conceptual Evidence Collection and Analysis Methodology for Android Devices

Ben Martini; Quang Do; Kim-Kwang Raymond Choo

Information Assurance Research Group, University of South Australia

ben.martini@unisa.edu.au; quang.do@mymail.unisa.edu.au; raymond.choo@unisa.edu.au

**Abstract**

Android devices continue to grow in popularity and capability meaning the need for a forensically sound evidence collection methodology for these devices also increases. This chapter proposes a methodology for evidence collection and analysis for Android devices that is, as far as practical, device agnostic. Android devices may contain a significant amount of evidential data that could be essential to a forensic practitioner in their investigations. However, the retrieval of this data requires that the practitioner understand and utilize techniques to analyze information collected from the device. The major contribution of this research is an in-depth evidence collection and analysis methodology for forensic practitioners.

## 1 INTRODUCTION

The Android operating system (OS), released to the public by Google in 2008 [1], is currently the most widely used smartphone OS, with a market share of 44% as of the first quarter of 2014 [2]. As the number of Android users rises, the potential for evidence of crimes (both cyber and traditional) to be stored on Android devices increases. This results in the need for a forensically sound methodology for extracting and analyzing evidential data from Android devices. Digital forensics is concerned with identifying, preserving, analyzing and ultimately presenting digital evidence to a court [3]. For Android data collection techniques to be considered forensically sound, they must follow the digital forensics principals of upholding data integrity, correctness and preservation. However, much of the existing research in this field (see Section 2.2) has made compromises in these respects with the aim of acquiring all possible data on the device.

Many types of potential evidential information can be extracted from a mobile device including SMS messages, phone call logs, photos and location data. While extraction of these types of data has been commonplace for some time, more recently a focus has been placed upon collecting data from third-party apps that users install on their mobile devices. Within the many categories of popular apps (e.g. games, shopping, finance), cloud computing apps seem some of the most likely to store data which could be of potential evidential interest in a criminal investigation or a civil litigation. Of the cloud computing apps on the Google Play Store, cloud storage and note taking apps are some of the most downloaded. However, there has not been a commensurate level of research conducted on the most effective method of collecting and analyzing evidence from cloud apps on Android mobile devices.

Much of the research which has been conducted in this area has also aged, as Android continues to advance and change with new versions being released on a regular basis. As such, we chose to focus on cloud apps installed on contemporary versions of Android as a case study for our methodology for collecting and analyzing app data for forensic purposes [4].

The structure of the remainder of this chapter is as follows: Section 2 discusses related work in this field, in particular, looking at existing techniques for the collection and analysis of forensic artifacts on Android devices. Section 3 describes our methodology for collection and analysis of evidential data on Android mobile devices. The methodology focuses on adherence to forensic soundness principles and aims to be as device agnostic as possible. Martini and Choo [5]'s published cloud forensics framework is used as the underlying guiding context for this research and as such the process stages are represented within the four stages of this framework (also see Figure 1). Finally in Section 4, we conclude our work with a summary of our findings.

## 2 RELATED WORK

### 2.1 BACKGROUND

The Android OS runs as middleware on top of a Linux kernel [6] meaning many Linux specific security features are used by Android. These security features include assigning each installed app a unique user ID and running each app within its own virtual machine [7], effectively sandboxing them. Android also implements a permission-based system to further enhance security on the device.

When an Android app is to be installed, it first requests a list of permissions that it requires which the user must then accept to install the app. These permissions include access to device resources such as Internet, Bluetooth and External Storage. If an app requires access to these resources, the developer must state so in the app's "AndroidManifest.xml" file located within the Application Package (APK) for the app. Android apps are generally written in the Java programming language and converted into a format (DEX) compatible with the virtual machine used in the Android OS, known as the Dalvik Virtual Machine (Dalvik). In addition to permissions that normal apps (apps that are installed into the "/data/app" directory) may use, there are system level permissions that may only be used by apps which are installed in the "/system/app" directory and signed with the developer key that was used to sign the device's OS. These apps are known as system apps. The "data" and "system" directories, along with several others, are mountpoints for partitions that Android uses in order to further compartmentalize the system.

A typical Android device has the following partitions:

- **system** – Contains system files, system apps, device frameworks
- **data** – Contains user installed apps, user data and Dalvik cache
- **cache** – Stores temporary system data (such as when installing apps)
- **boot** – Boot partition, used for normal booting of the Android device
- **recovery** – Recovery mode partition which the device can alternatively boot to

The Android Debug Bridge (ADB) is an application designed by Google, which comes packaged with the Android SDK, to allow programmers to perform a range of useful tasks related to debugging Android apps. ADB allows a developer to connect an Android device (with ADB debugging enabled) to their PC and debug it via USB (or TCP). For example, ADB is capable of directly installing apps onto the device, bypassing any installation screens and furthermore able to retrieve files from the device directly to the connected PC. A developer can even open a command line interface (shell) into the Android device and perform further tasks. In order to perform higher privileged tasks such as rewriting files on the "system" partition, the Android device typically must be rooted.

Rooting is the process of altering (usually exploiting) a device in order to gain administrator or "root" privileges. The process of rooting an un-rooted device generally requires the use of a technique that exploits a security flaw on that particular Android device and OS build. This is because device manufacturers generally do not release their devices with root access enabled. Due to the software and hardware specific nature of these exploits, rooting a device to obtain forensic data is not ideal. Furthermore, root exploits are generally not open source (due to the possibility of a manufacturer releasing a patch) and, therefore, a forensic practitioner may not be able to fully explain what the exploit may have done to the device. The process of rooting a device may also change data on the device in important partitions (such as "system" and "userdata"). Due to these considerations, rooting should generally be avoided in a forensic collection process.

## 2.2 EXISTING LITERATURE

For the purposes of this research, we consider techniques, methods and approaches to have the same definitions and to mean a singular way of performing a task or achieving a goal. These are very low level and go into specific details (e.g. a method for obtaining a "bit-for-bit" copy of a partition from an Android phone). Models, frameworks and processes are considered in this research to be high level abstract forms of techniques, methods and approaches. Instead of details, these are a series of operations to be used as the backbone or as support for techniques, methods and approaches. They are often used in order to bring together a series of techniques into one cohesive piece. An example is a process which can be followed that collects data of forensic interest from an Android device. We consider a combination of a number of high level, low level or both levels of procedures to form a methodology.

Current digital forensics techniques for extracting data from an Android device can be categorized into:

- live analysis where the forensic information is taken directly from the device, and
- offline analysis where a copy of the device's data is analyzed.

Maus, Höfken and Schuba [8] proposed an approach for obtaining and analyzing location data on Android devices. The authors performed an extraction of all the data on a rooted Android device and then analyzed the resulting data for databases containing certain attributes (such as "longitude") and photos with EXIF geo-data. The location information obtained is then plotted onto a map with timestamps and the estimated routes. Similarly, Spreitzenbarth, Schmitt and Freiling [9] investigated the sources for location data on Android devices and found that many system apps stored a large amount of location data within their caches. The methods utilized by the authors in order to obtain the user data required for analysis once again required the device to have root access.

Al Mutawa, Baggili and Marrington [10] proposed a method to analyze several smartphones for evidentiary data in the form of social networking app artifacts. Once again, the authors were required to root the phone in order to access the user data needed to create a logical copy of the phone. The authors were able to recover information such as usernames, passwords, profile pictures and chat messages.

Kim, Park, Lee and Lee [11] proposed a technique to analyze a user's smartphone usage behavior by taking advantage of ext4 journal logs used by the Android OS to provide fault tolerance. They obtained a logical copy of the data partition of the Android device by rooting it and running the "dd" command. By using the journal logs, the authors were able to find out what actions a user performed (such as accessing or deleting a file) and recover certain deleted files.

Meanwhile, research by Barghouthy, Marrington and Baggili [12] attempted to retrieve private browsing session information from an Android device. Without rooting the device, the researchers found that no private web browsing history was accessible. After rooting the device, the researchers were able to find the user's browsing history and login names.

Research by Andriotis, Oikonomou and Tryfonas [13] shows that while it is possible to obtain information such as files transferred via Bluetooth and visited wireless networks, this information is stored in a very small buffer. This means that the information obtained depends heavily on the time of the seizure of the device. To access this information (and to obtain a logical copy of it) also requires the device be rooted.

Finally, Chung, Park, Lee and Kang [14] focused on extracting useful information off smartphones pertaining to apps and app data. They found that in order to access useful information such as app access keys and secret keys, the device was required to be rooted. Files such as those downloaded by cloud apps were generally kept on the external SD card and could be accessed without rooting the device.

It is clear from the literature review that existing research does not use a forensic framework as the basis of their evidence collection and/or analysis techniques. To contribute towards filling the literature gap, we will describe our proposed evidence collection and analysis methodology for Android devices in the next section.

# 3 AN EVIDENCE COLLECTION AND ANALYSIS METHODOLOGY FOR ANDROID DEVICES

This section outlines the methodology that we have developed for collection and analysis of evidential data from modern Android mobile devices. The methodology is constructed using the principles of Martini and Choo's cloud forensics framework [22] to ensure forensic soundness.

Mobile devices (vis-à-vis 'traditional' mobile phones), having relatively recently gained maturity as a technology, tend to natively employ security features that standard PC OSs are only now beginning to introduce. Android devices are no exception to this with advanced app sandboxing to ensure that user privacy is preserved, device encryption and increasingly advanced login systems based on a range of technologies varying from simple and complex passcodes through to biometrics. This presents a far greater challenge to a forensic practitioner in comparison to traditional PC based forensics.

When designing the evidence collection and analysis methodology, we sought to implement two key characteristics:

- **Adherence to forensic soundness principles** – One of the key characteristics of the process was to maintain forensic soundness especially in terms of data handling. The first two 'rules' which McKemmish [3, p.3] notes in his seminal paper on digital forensics are *'Minimal Handling of the Original'* and *'Account for Any Change'*. In the context of contemporary smart mobile devices, this can be interpreted as ensuring that the absolute minimum changes required are made to the mobile device to collect its evidence.

  In most cases, it would be unrealistic to assume that it is possible to collect all potential evidence from a mobile device in a timely and accessible manner without at least minor modifications (e.g. bypassing security features on the device). However, adherence to McKemmish's second rule means that any operation that a practitioner undertakes on a mobile device must be fully understood by the practitioner. This is to ensure that the minimum number of changes are being made to the device, that the integrity of the evidence on the device remains intact, and that the precise steps undertaken on the device can be explained to a court as part of the reporting and presentation process.

To date, a number of the published digital forensic processes for Android devices use methods that involve unknown instructions being executed on the device (as discussed in Section 2.1). This is most commonly part of the 'rooting' process or flashing a third-party (unverified) 'recovery OS'. A number of authors [8, 10-13, 15, 16] use tools developed by third parties that are generally not designed for forensic collection purposes and, as such, may result in the integrity of the evidence collected being questioned as part of legal processes (e.g. changes to user data partition without the forensic practitioner's knowledge).

- **Ensure the process is device agnostic as far as practical** – Another characteristic we sought to implement when designing the process was to ensure that the process was as device agnostic as possible. This is not a straightforward aim as the Android OS is known for the fact that it runs on many different devices manufactured by different organizations. Each of these parties almost always customizes the OS resulting in what is known as 'fragmentation' or 'fracturing'.

  From a research perspective, this means that all processes designed for Android forensics generally apply only to specific devices and particular OS versions. Our process does not aim to resolve this issue entirely; due to the customization available to OEMs and carriers, this may be an impractical aim. However, we aimed to document methods (wherever possible) that operate on the underlying Android OS and APIs rather than relying on, for example, exploits in particular customized OS libraries or utilities. This characteristic must be rationalized against our requirement for forensic soundness which is the overriding factor in the design of any digital forensic process.

Our proposed collection methodology comprises eight steps, as outlined (in bolded boxes) in Figure 1, which we discuss in the context of the cloud forensic framework [5] to provide background.

The methodology includes techniques to bypass device/OS security features, collect a forensic "bit-for-bit" image of the device's data partition, analyze the collected image and where necessary inject the OS libraries with code to retrieve securely stored data and/or credentials.

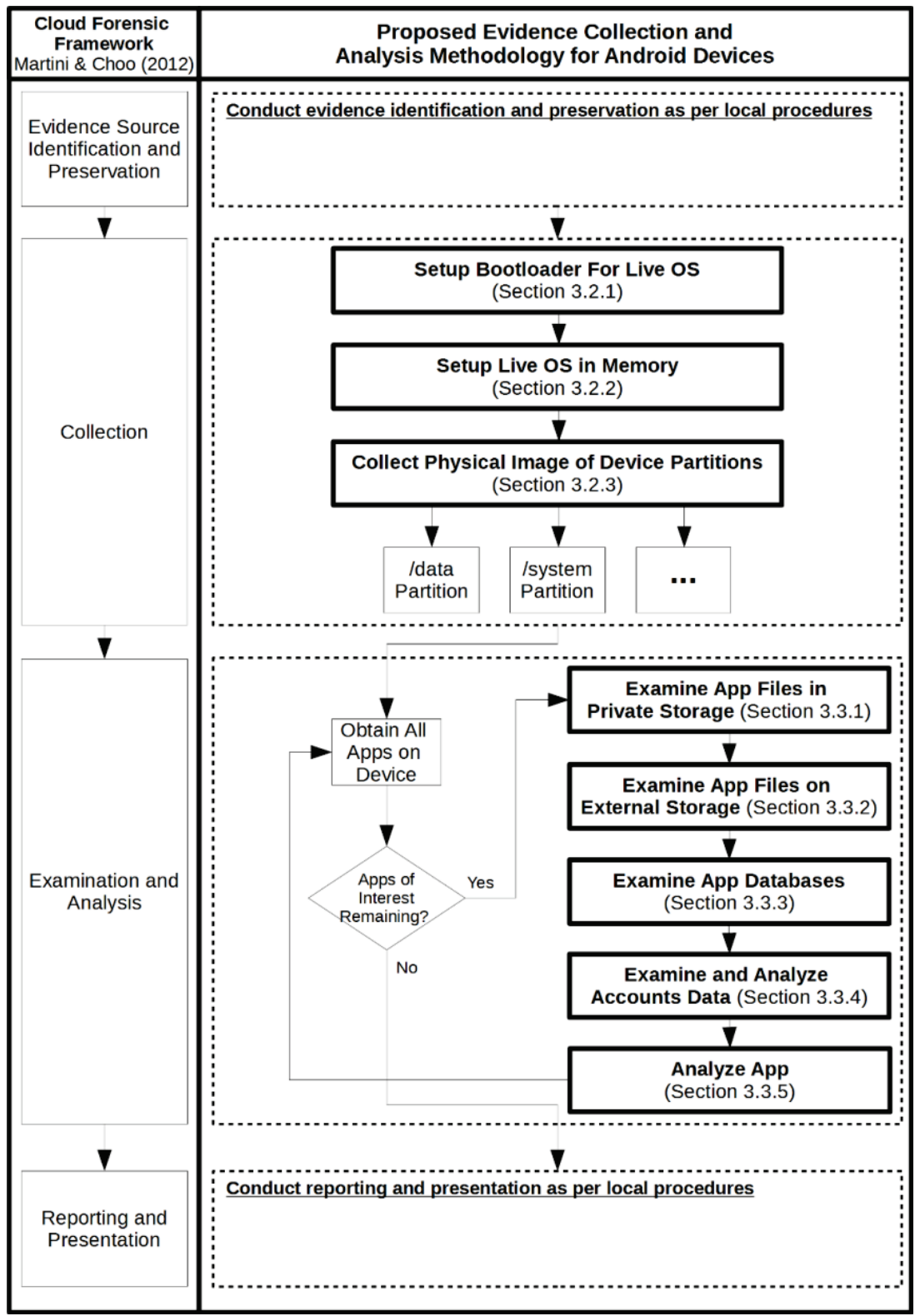

*Figure 1 Evidence Collection and Analysis Methodology for Android Devices*

## 3.1 IDENTIFY DEVICE AND PRESERVE EVIDENCE

The first step in our methodology for evidence collection is to identify the device and approximate the Android version so that we can use device or OS version specific methods, where necessary, for the remaining stages of the technique. This assumes that the practitioner has already placed the device in a forensic Faraday bag. Otherwise, this should be completed as a first step when the device is seized.

It is important that the device remains in a radio suppressed environment (e.g. the use of a forensic Faraday bag) while it is powered on. This is necessary as it is common for devices to have the ability to be 'remotely wiped' when sent a signal over a mobile or Wi-Fi network. If this occurs, it will result in the device erasing its internal (and, potentially, SD card) memory. In many cases, this will prevent any further forensic analysis and, as such, should be avoided at all costs. If it is not possible for the device to remain in a radio suppressed physical environment and the device is unlocked, practitioners may choose to place the device in "Airplane Mode" which is designed to disable the device's radios (for use on airplanes). While this is advisable in comparison to leaving the device powered on with access to a mobile network, it introduces unnecessary risk in comparison to using a physically radio suppressed environment. Practitioners should also follow standard practices in terms of physical evidence preservation, particularly in terms of ensuring that they take appropriate photographs of the device and any attachments when it is seized.

If the device is unlocked when initially seized, the practitioner should confirm if device encryption is enabled and/or if a lock code is enabled. If device encryption is enabled, then the device should not be powered off as this will require the encryption password to be entered before the data on the phone can be accessed. At the time of research, there is no supported method for disabling device encryption without wiping the contents of the phone's data partition. As such, if device encryption has been enabled, the practitioner should consider conducting a logical acquisition using methods such as Android device backup. This may also be a prudent step (if the device is unlocked) regardless as additional evidence collection methods should be used where feasible. Although a logical collection will not be able to provide all the data possible via physical collection, it does offer a useful backup. Practitioners may also consider taking screenshots of the state of the device (including any virtual connections, e.g. to wireless networks) when the device is seized if it is unlocked. However the benefits of this approach must be balanced against the risk of the screenshots overwriting potential evidence that has been deleted and is located in unallocated space.

If a practitioner finds that a lock code is enabled on an unlocked device, the practitioner should consider disabling the lock code (a change which can be accounted for). While the device is powered on, the practitioner should attempt to determine the version of Android that the device is running. If the device is unlocked, this information can be gathered from the "About Phone" section of the Settings app, and the device model can also be obtained from this section. If the device is locked, the practitioner may be able to use visual artifacts on the lock screen such as the user interface style to provide an indication of the approximate Android version.

Recently, there have been a number of publications on the topic of forensic collection of volatile memory in Android devices (see [15, 17, 18]). It may be necessary to collect the volatile memory of the device, as a form of preservation, before the primary evidence collection stage to ensure that potential evidence is not lost when the device is powered off. This is particularly pertinent when the device is encrypted as the encryption key may be stored in volatile memory. Unfortunately all of the techniques that we reviewed (for modern versions of Android) required that the device be rooted. This is due to Android's sandboxing model preventing apps from reading the memory space (or heap) of another app. If the device is already rooted and unlocked, then this collection method may be an acceptable forensic procedure. However, the vast majority of Android devices will not be rooted. Many authors do not mention the requirement to root the device for forensic collection and those who do generally do not discuss the forensic implications of running code that executes unknown instructions on a device undergoing forensic analysis. As rooting relies on exploits which are generally patched in future versions of Android, their success cannot be guaranteed. Many (if not most) root exploits for modern Android mobile devices also require that the device be rebooted, which negates their use for volatile memory collection.

Once these steps have been completed the device can be powered off. After the device has been powered off, the next step is to identify the device, which is generally achieved by looking for manufacturer, model and other device specific markings or labels on the device (generally on the back cover).

## 3.2 COLLECT EVIDENCE

Once the device has been identified and preserved, the next step is to collect a physical forensic image of the device's 'userdata' partition. This partition is generally non-volatile, as such practitioners should have preserved (and collected) all volatile data by this point (as discussed in Section 3.1). Under normal circumstances (as enforced by Android security), the "userdata" partition is the only internal partition to which a user can directly write data and, as such, is the most likely location for evidence to be stored. However, if the practitioner has any doubts as to the integrity of the Android security on the device (e.g. if the device is suspected of being rooted or is running a custom OS), they may choose to collect physical images of the other partitions on the device (see Section 2.1 for a list of common Android partitions).

Physical collection is preferable to the more common logical collection (via the stock OS) for a number of reasons. Physical collection allows for a more thorough analysis of the data stored on the device's flash memory by ensuring that we collect all of the files on the partition. In comparison, a logical collection using a tool such as Android backup will only return the files that the OS provides as part of a backup (this tends to be somewhat vendor specific, but is generally a subset of the available data). The collection of a physical image also allows for the potential recovery of deleted files from unallocated space, file signature verification and header/keyword search using existing forensic tools that are able to read the ext4 partition format used by modern versions of Android.

Another major advantage of physical collection is that it will circumvent passcode lock controls on the device by bypassing the stock OS, something that will generally thwart a collection process that requires the device to be rooted and uses the stock OS.

When conducting forensic analysis on a traditional PC, a common technique for physical collection is to boot the PC using a triage live OS (which runs in RAM) or to remove the disk(s) from the PC and use a write blocker in a second PC to collect an unmodified physical image. These techniques have parallels in collection of evidence from Android mobile devices. The physical disk collection method is analogous to the 'chip-off' (i.e. physically removing the NAND flash from a device) technique used in mobile forensics. While this method has proved useful, it is very time consuming and the technique may not be perfected for the flash storage used in each type of Android mobile device. The live OS collection method is used for physical collection of other mobile devices (e.g. the custom ramdisk approach used in 'iphone-dataprotection'[19]) and can be applied on any Android device that is capable of booting from a PC via USB or external media (e.g. SD card). This method has many advantages including the speed and effort required for the collection and the relatively device agnostic process in comparison to chip-off analysis.

The selection of live OS to run on the device is integral. Wherever possible, practitioners should build an appropriate live OS using the AOSP and vendor supplied source code/drivers. Where this is not feasible the practitioner should only use a prebuilt live OS when it was designed for forensic extraction and the source of the OS is trusted.

Our technique for booting the OS into RAM on the mobile device is discussed in Sections 3.2.1 – 3.2.2.

### 3.2.1 Setup Bootloader for Live OS

The first step in the collection stage of our technique is to modify the bootloader so that it will allow us to boot a live OS. While in some cases this may be possible with the default factory configuration, most devices will have some security applied to the bootloader preventing the loading of custom OSs. One prominent reason for this is that manufacturers generally terminate support and warranty arrangements when a user installs a custom OS, as such they configure their devices to warn users of this and may require that they retrieve a device specific 'key' from the manufacturer's website. Some manufacturers also note the security flaw we are seeking to exploit, in this case, for forensic collection. By running a custom OS, we can bypass the security enforced by the stock Android OS (e.g. sandboxing and file system/device permissions) and read from the flash memory/secure storage/etc. directly.

The process for reconfiguring the bootloader to allow us to boot a live OS is commonly known as "unlocking" a bootloader. This process is device-specific (hence the requirement to gather device information as part of the first stage of this technique) although a number of manufactures use similar processes that make use of the "fastboot" tool. Many modern Android devices can be placed in "fastboot mode" (via a key combination while the device is powered off) that will allow us to provide instructions to the bootloader using the fastboot tool. Once the device is in fastboot mode, unlocking the device may be as simple as passing the "oem unlock" command via the fastboot tool. The device may respond advising that the unlock procedure has been successful or may require (as discussed above) that a unique string (displayed at this point) be provided to the manufacturer to retrieve an unlock key which will need to be provided at this point. Generally, the practitioner will be able to retrieve this key via the standard web form used by customers for this purpose.

Some manufactures may choose to wipe the Android device's data partition as part of the unlock process. If allowed, this will prevent us from accessing the data on the mobile device and potentially destroy it, as such the wiping process (if it exists) must be bypassed or disabled. This will require research on the part of the practitioner to determine an appropriate exploit to bypass this process. This may be somewhat time consuming unless this research has already been completed and published by a trusted source. Where wiping (or any destructive operation) is involved, a practitioner is strongly advised to obtain an identical device to that which is being analyzed, which they may test the technique on before attempting to apply it in the investigation.

### 3.2.2 Boot the Live OS in Memory

Once the bootloader has been unlocked, we can commence booting the custom live OS (i.e. the custom image). In the majority of cases, a practitioner or their organization should develop their own live OS based on the Android core (kernel and low level services).

Votipka, Vidas and Christin [20] described a method for creating recovery images for forensic collection purposes. This method can be adapted to create the live OS image that is similar to a minimalist recovery image. Creating a live OS image is advantageous in a number of ways but primarily it provides the practitioner (or organization) with a higher level of confidence that the processes being run on the device are forensically sound (e.g. ensuring that partitions on the device are not mounted until an image has been collected). Creating the OS also ensures that the practitioner is aware of, and, as such, can account in their reporting for, any changes that are required to load the live OS on a particular device.

Each device will require a specific kernel and/or drivers and, as a result, each live OS image will require customization on a per device basis. However, once it is developed, it should not need to be updated for each individual device. Where it is infeasible for the organization or practitioner to create the live OS, it may be acceptable for the practitioner to use a prebuilt live OS designed for forensic collection (assuming the practitioner trusts the source of the OS). It is strongly recommended that live OS images that are not specifically designed for forensic collection not be used for this purpose.

Apart from the potential for malicious intent, images may unintentionally erase or modify data that may be integral to a forensic case and in a manner that is unpredictable.

Once the image has been built, the next step is to boot the image into the volatile memory of the device. Notably this ensures that no data needs to be wiped on the device as would be required if an OS image was flashed onto the device. A number of existing Android forensic processes flash the "recovery" partition on the basis that user data cannot be stored there (see [20-22]). However, this is not universally true. If this becomes a common procedure for forensic collection, then suspects can begin hiding data in this partition that will implicitly be erased when the forensic collection image is written to the partition. This is analogous to writing a triage OS partition to the disk of a PC so that we can boot the forensic tool that will implicitly delete any data (presently allocated or unallocated) that was previously written to that location on disk.

Again using the popular fastboot system as an example, once the OS image has been created, booting the image into memory is as simple as issuing the "fastboot boot liveos.img" command from the PC to which the mobile device is attached. The live OS image file will be transferred to the mobile device memory and the device will attempt to boot it. Once the OS has completed booting and the ADB daemon has started, we can commence the final stage of image collection.

### 3.2.3 Collect the Physical Image of the Device Partitions

The forensic copy technique used for physical collection of the partitions and transfer of the "bit-for-bit" images to the PC will depend on the method the practitioner has chosen to implement in their live OS. The practitioner can set up this process to suit their organizational/legislative requirements. We propose using a generic process for physical collection that makes use of a common Android toolset while ensuring forensic soundness, such as the following:

ADB supports forwarding of various sockets/devices to a PC via USB. We use the "adb forward" command to forward the mobile device flash block devices to the PC via USB for collection. This is achieved by using the ADB application on the pc running a command such as "adb forward tcp:7000 dev:/dev/block/platform/[flash memory name, e.g. "msm_sdcc.1"]/by-name/userdata" which would allow the practitioner to access a "bit-for-bit" stream of the "userdata" partition on the phone by connecting to TCP port 7000 on the host PC. In practice this is achieved by having an application read from port 7000 until the end of stream and writing the received bytes to a binary file. The collection of the "userdata" partition will also result in the collection of all data on an emulated SD card (located at "/data/media/0" within the "userdata" partition). In the case of a physical SD card, the practitioner will need to undertake a separate physical collection. If they choose to do so via the live OS, it can be accessed at the "/dev/block/platform/[SD card memory name, e.g. "s3c-**sdhci**.2"]/mmcblk1" path and collected using a similar technique to that used to collect the "userdata" partition.

Once the collection has been completed, the practitioner should use one or more hashing algorithms to verify the integrity of the collected image. Using the 'md5sum' and 'sha1sum' tools on the mobile device's flash block device file and on the collected image files stored on the PC should result in identical hash values. These hash values should be recorded for later use as part of the reporting stage.

## 3.3 EXAMINATION AND ANALYSIS

Using the physical images collected from the mobile device, we can now begin examination in order to extract any potential evidential data. We will also use the information gathered from the examination stage to guide the types of analysis that will need to be conducted on the physical images and, potentially, the mobile device to retrieve further evidential data.

The first step of examination begins with determining the installed apps and which apps are of particular interest to this examination. While this stage does not prevent the practitioner from using standard forensic techniques on the physical image, such as a tool to conduct a raw keyword search for known keywords of interest, it allows us to focus the examination on and deliver in-depth results for particular apps. The apps of interest may be of a particular category (e.g. cloud apps or communication apps) or may be individual apps depending on the nature of the case.

There are a number of methods for determining the apps that have been installed on an Android device. However, once a physical image has been collected, one of the most straightforward methods is simply to list the subdirectories in the "data" directory of the "userdata" partition. Each of these subdirectories will have an app "package name" (e.g. "com.example.appname"), notably this name is not the friendly name which is used to display the app in the OS, but rather it is a unique name selected by the developer conforming to a certain standard. Generally the app friendly name will be part of this app name. This subdirectory is known as its data directory or private app storage area.

Private storage is commonly the main focus for a forensic investigation on a particular app, as Android uses sandboxing to prevent apps from interfering with each other and the OS. This also means that apps are restricted to writing files to their private app directory (and selected other places such as external storage). An app can write almost any type of file to its private storage, although it is generally preferable to write large files to "external storage" (often a physical or emulated SD card), as private storage is often limited. Private storage commonly stores the following types of files:

- "webview" data (such as cache and cookies),
- databases (generally SQLite),
- files (can be any type of file),
- cache (app cache data),
- lib (app software library files) and
- shared_prefs (preference files in a common Android XML based format).

Once the available data has been examined (and if necessary analyzed) from both private and external app storage, we may need to conduct further analysis operations on the mobile device to collect any other potential evidential data or data required to further the collection with other evidence sources beyond the individual mobile device being analyzed. This would be conducted as a separate iteration of the forensic framework (as outlined in [5]).

Our technique for examining the available evidence stored by apps and conducting further analysis on an Android device is discussed in the following five subsections.

### 3.3.1 Examine App Files in Private Storage

Starting the examination with each app's private storage directory seems to be the logical choice as the majority of app specific data is likely to be stored there. There are no mandatory items that need to be stored in private app storage and, as such, the exact data that will be extracted will be app dependent. However, we aim in this section to provide an overview of the types of data that would be commonly found in private app storage.

Configuration files known as "shared_prefs" are commonly used in Android apps. Access to these shared_prefs files is abstracted in the Android API, simplifying configuration directive storage and making it easily accessible to every app developer. Developers can store a wide range of data in shared_prefs files that are located in the "shared_prefs" subdirectory of the app's private storage directory.

Practitioners should ensure that they examine the contents of these shared_prefs configuration files to not only locate potential items of evidential interest (e.g. app event timestamps, credentials, user identification information, recent items, etc.) but also information which may aid in the further examination and analysis of the seized device (e.g. encryption keys, database locations) and of the user's other devices (e.g. remote server/cloud information).

Two formats are commonly found in shared_prefs files: An XML structure for standard objects such as strings, integers and Booleans and, within the XML, JSON encoding is commonly used to represent arrays of objects that have been serialized from the app. While the meaning of configuration directives in these files is not always obvious, it is generally in plain text and a meaning can often be inferred. Where there is ambiguity over the meaning of the configuration directive, a practitioner can use analysis techniques to attempt to determine a certain meaning, this is discussed further in Section 3.3.5.

Following on from the analysis of the app's shared_prefs, the practitioner should examine the other files stored by the app for evidential value. Application libraries (stored in lib) are unlikely to be of value to most forensic investigations. Cache files, if present, may expose evidential data that was temporarily stored by the app; however non-standard binary formats are commonly used. Unless the format can be decoded, binary analysis of the strings may be the only straightforward means for analysis. Other files, including those stored in the 'files' subdirectory, may be in any format of the app developer's choosing. Standard forensic techniques, such as header analysis, can be used to determine the file type and potentially decode the file.

Web view data (cache, cookies, etc.) is stored in the "webview" subdirectory when web views are used in the app. Analysis of this information, in a similar manner to traditional browser analysis, may provide evidence of interest.

Databases are commonly used by apps to store a range of potential evidential data including configuration information entered by the user, metadata stored by the app (especially reflecting app usage) and even the data stored by the user (depending on the nature of the app). App databases are discussed further in Section 3.3.3.

### 3.3.2 Examine App Files on External Storage

Following the examination of the files stored by particular apps within their private storage directory, a practitioner should examine files on the device's external storage. The name "external storage" is somewhat ambiguous; but for the purpose of this research, generally refers to storage on the device either provided via a physical SD card inserted into the device, emulated SD card (where data is stored on the phone's internal flash) or a combination of both. Regardless of the type, external storage is one location on the device where Android's sandboxing rules are significantly relaxed.

Generally any app that has been granted the read and/or write external storage permission can read and/or write to any part of the external storage (unless a third party solution, such as the one described by Do, Martini and Choo [23], is protecting the storage). This means that an app can write data to any location on external storage. However, there is a special directory on external storage specifically structured for individual app storage. It is similar to private storage but does not have the sandboxing protection available in private storage. Apps that make use of this facility will have a directory with their app name (identical to their subdirectory within the "userdata" partition's "data" directory) in the "Android/data/" directory of the external storage. Depending on directory permissions, it may be reasonable to presume that data stored in these directories was created by the app in the directory name.

The external storage should be examined in a similar fashion to the private app storage with a similar range of files being potentially stored on external storage. As noted, larger files are likely to be stored on external storage; however, any type of file can potentially be stored on external storage. Due to the lack of security controls, some apps may opt to encrypt their files which are stored on external storage.

### 3.3.3 Examine App Databases

As both private and external storage are being examined (particularly the private app storage databases directory, but potentially in any app-writable directory on the device), a practitioner will likely locate and collect various databases of interest. SQLite is natively supported by the Android API and as such is commonly used. However, this does not prevent apps from using other database formats. For example, apps may choose to store data in a proprietary format and/or a format more suitable to their data (e.g. a key-value pair database for efficiency).

Header analysis can generally be used to identify the type of database and allow the practitioner to locate tools suitable for decoding the database. This may not be the case if the entire database file is encrypted by the app, in which case the practitioner will need to locate the database and key based on configuration file data (see Section 3.3.1) or further app analysis (see Section 3.3.5).

Similar to shared_prefs or other configuration files, the meaning of the columns and tables in a RDBMS type database, for example, should be somewhat self-explanatory as the original descriptive strings should still be available. However, in the case of ambiguity, further app based analysis can be completed as discussed in Section 3.3.5.

### 3.3.4 Examine and Analyze Accounts Data

For apps that use online services requiring authentication, the next stage is to examine and, if necessary, use analysis techniques to collect the accounts data stored by Android. Apps may choose to store their accounts (mostly credentials) data in a secure database provided by the OS AccountManager service. The AccountManager API has been designed to ensure that only the app that stored a particular set of credentials is able to retrieve them. In our experiments using seven popular cloud apps described in [4], we found that there are at least two ways for this accounts data to be stored on the device. The data can be stored in an SQLite format database located at "system/users/[device user id, generally 0]/accounts.db" on the "userdata" partition. Alternatively, if implemented by the device manufacturer, accounts data can be stored in a physical secure store on the device SoC.

In the former case, analysis of the SQLite database (which was collected as part of the physical image of the "userdata" partition) is quite straightforward. In our experiments described in [4] we found that the tables of particular interest were "accounts", "authtokens" and "extras", although the practitioner should examine all tables for potential evidence.

The accounts table is used to store account records including their "name", "type" and a "password" which may contain: a password, a string other than a password (e.g. an authtoken) or be blank.

The authtokens records contain an "accounts_id" (to link the authtoken to their account record), a "type" and the "authtoken" (a string defined by the app or authentication service provider).

The extras table stores data added to secure storage using a key-value system which requires that the calling app not only be the app which created the associated account, but also that it has the relevant key to retrieve the value (retrieval requests with unused keys return null). This table could store a range of data but commonly includes refresh tokens, account IDs, etc. The fields in the table include "account_id", "key" and "value".

If the device does not use this SQLite database and instead chooses to store the credentials in physical secure storage, this complicates collection of these credentials. However, in cases where evidence collected from remote services is integral (such as those involving cloud computing or social networking), we may need to conduct further analysis to retrieve these credentials from the device.

The analysis technique that we propose requires changes to be made to the device's software that would require 'handling the original'. We believe that this discrete change, which the practitioner will be able to account for, should not be able to cause modification to the evidence. As with any change, a practitioner or researcher should weigh the cost of making the change against the potential to collect further evidence and make a decision on whether to proceed on a case-by-case basis.

To proceed with the necessary changes, the practitioner will need to once again boot the phone into the live OS (as discussed in Section 3.2). They will then need to mount the "system" partition and extract the "services.jar" (and potentially "services.odex") file from the "framework" directory. If the "services.jar" which is included on the device does not contain a "classes.dex" file when decompressed from the ZIP format, then the practitioner will need to extract and deodex the "services.odex" file, otherwise they will need to "baksmali" the "classes.dex" file. In either event, the practitioner will generate a set of SMALI source files for the Android services framework.

"smali" and "baksmali" (see https://code.google.com/p/smali/) are an assembler and disassembler (respectively) for Android apps.

- baksmali is used to convert an Android app's "classes.dex" file into human readable SMALI code.
- smali is used to convert these SMALI files back into a DEX format file that is compatible with the Android Dalvik virtual machine.

The practitioner should locate the "com/android/server/pm/PackageManagerService.smali" file within this source tree and then modify and inject the "compareSignatures" method (which returns a Boolean) such that it always returns "true" when receiving the practitioner's package signing key. Injecting the "compareSignatures" method such that it always returns "true" is a known technique discussed in the Android development community [24].

The SMALI source tree should then be recompiled (using the "smali" application), the "classes.dex" file should be replaced in the "services.jar" compressed file and the updated jar file should be uploaded via the live OS to replace the original on the device's system partition. The original "services.odex" file on the device is deleted in order to trigger the OS to rebuild its system packages.

If the practitioner was required to de-odex the "services.odex" file to obtain the SMALI code, other system packages may require de-odexing as the Android OS detects all changes in system files and attempts to rebuild the package and ones it is dependent on. As the other JAR files do not contain a "classes.dex" file for Android to turn into an "odex" file, the OS may not boot.

We developed a script to automatically rebuild system JAR packages with their respective "classes.dex" files by monitoring the "logcat" output (the development log which is visible from a connected PC). Any dependencies that cannot be rebuilt are logged in the "logcat" output as a "StaleDexCacheError" type message containing the required package's name. This script simply parses the output for these strings (which contains the system package required to be de-odexed). This process is outlined in Figure 2.

Once this replacement is completed, a practitioner can develop and sign an APK which requests the relevant account credentials from the device's secure storage at boot and outputs the credentials to "logcat". This is achieved using the methods that are available when instantiating the AccountManager class. For example, an array containing all accounts on the device can be obtained with the AccountManager "getAccounts" method. Using this, the AccountManager "getPassword" method can be passed an "Account", which will return the "password" (plaintext password, authentication token, etc. depending on what the app stores) for that particular account. The "getPassword" method ordinarily requires the signature of the caller to match that of the account being obtained but the method described above gives this APK effectively system level permissions and, therefore, permission to read all data from all accounts stored upon the device.

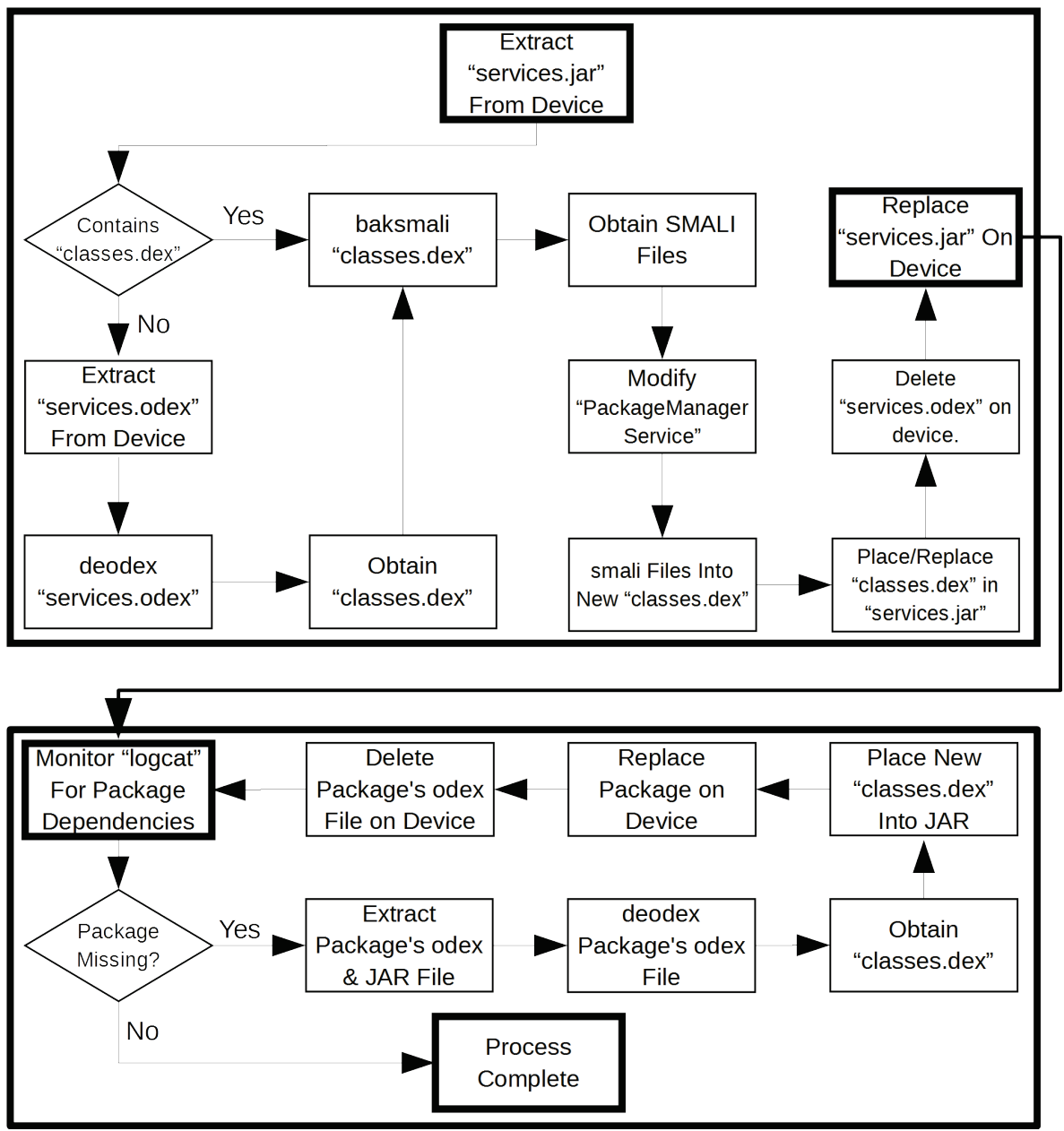

*Figure 2 Modifying the "services.jar" File (Top) and Package Monitoring (Bottom)*

Notably this technique will work on devices with secure storage and, with some modification, devices where the user has enabled encryption. This is possible as Android does not encrypt the system partition when device encryption is enabled.

### 3.3.5 Analyze Apps

Once all of the above data has been examined, and analyzed where necessary, the practitioner will have collected the majority of the body of potential evidence on the device. However, there may be some remaining ambiguity at this stage as to the exact meaning or source of some of the evidence. Before reporting and presentation commences, the methodology aims to reduce, as far as practicable, any ambiguity as to the provenance or meaning of the collected data. As such, the final stage of analysis and examination is to analyze the underlying code of the apps being examined.

There are a number of approaches to this stage of the technique which, when combined, can assist a practitioner in efficiently gathering the background information on the located evidential data. The approaches selected will vary depending on the case and apps being analyzed. Broadly speaking, most authors refer to two types of app analysis: static analysis where the app code/binaries are analyzed to determine app operation, and dynamic analysis where the app is run to understand its operation. We propose a combination of both static and dynamic analysis approaches. This allows a practitioner to gain a complete understanding of the app while avoiding the shortcomings of the individual approaches (e.g. in the case of code obfuscation).

For most apps, it will not be necessary to analyze the original instance of the app on the evidence source device to determine its operation. The APK files for apps requiring further analysis can be copied from the "/data/app" directory on the "userdata" partition (or "/system/app" on the "system" partition in the case of system apps). Once collected, these APKs can then be installed on a development device or emulator that will allow the practitioner to have a greater level of visibility into the app's operations without requiring modifications to the original evidence source device.

Most approaches for analysis of Android apps are designed from the perspective of security researchers rather than forensic practitioners. Practitioners should investigate the contemporary analysis approaches available at the time of their investigation. However, as part of our proof of concept discussed in [4] we found the following techniques to be most useful for a forensic investigation.

One approach that we made extensive use of was analysis of app memory "heaps". These memory dumps represent the majority of objects that an app has currently provisioned in memory. Use of this technique on publically released apps (i.e. without a "debuggable" flag in their manifest) requires a development or engineering build of Android. We recommend that this build be installed in an emulator or virtual machine. Debugging will need to be enabled in the virtual Android device. Once debugging is enabled, the practitioner can attach the mobile device to a PC via the ADB (Android Debug Bridge) program and use the DDMS (Dalvik Debug Monitor Server) tool available from the Google Android SDK. DDMS supports the collection of app memory heaps in the HPROF (Heap PROFile) format to collect the current memory state of running apps.

We were able to develop an application for execution on the PC that will collect the memory heap of a particular app at regular time intervals (as set by the user). This allows a practitioner to examine the changes in the app's memory as it progresses through a procedure (e.g. during a login process). We used Eclipse Memory Analyzer to analyze the "HPROF" files, which requires that the Android "HPROF" file be converted to a generic Java HPROF format. This can be achieved using the "hprof-conv" tool available with the Android SDK. Eclipse Memory Analyzer initially loads a converted HPROF file into a view known as the Dominator Tree. In this view, one is able to sort all objects within memory by shallow heap (or retained heap). This is useful as larger objects contain more data which may be of use. The Dominator List can be further defined by filtering it to only show objects part of a certain class path (e.g. "com.dropbox.android"). In order to further analyze the memory heap, we must use Object Query Language.

Object Query Language or OQL is a query language designed for databases containing objects (i.e. object-oriented databases). The syntax of OQL is very similar to SQL with keywords such as "SELECT", "FROM" and "WHERE". Examples of tasks we can accomplish by executing OQL on the memory heap include:

- Obtaining a list of all Strings in the memory heap containing the word "authentication"
- Finding all instances of the "org.apache.http.client" class.

Decompiling the app and conducting static analysis of the source code also provided a useful approach to understanding the data stored by particular apps. A number of authors describe this approach (see [25-28]). The feasibility of this approach will be significantly reduced if the app code has been obfuscated (i.e. all identifying strings and notations – e.g. variable and method names – have been removed from the app). We found that obfuscation had been used in a number of the contemporary apps that we analyzed in [4]. While it is still possible to gain an understanding from code in this state, it becomes more difficult.

Analyzing the app's code is useful to determine how configuration entries are created, how the app databases are populated and read, and how security (such as tokens and encryption) is implemented and used.

## 3.4 REPORTING AND PRESENTATION

After the practitioner has completed examination and analysis of the relevant apps and their constituent data, they will need to produce reports and ultimately present their findings.

To demonstrate the integrity of the evidence that has been collected, cryptographic hashes (e.g. MD5 and/or SHA1) of the physical partitions (as taken from the block devices) and of the forensic images should match and be reported. This is standard practice for forensic image collection of PC hard disks and other traditional media; however, it has often been lacking in forensic analysis of mobile devices to date. This is presumably due to the difficulty in collecting this information and/or the difficulty in collecting a "bit-for-bit" image of the device's flash memory. This integrity verification process cannot be undertaken until the live OS has been booted, which demonstrates the necessity for the practitioner to completely understand the operation of the OS that they use. This is particularly notable as any minor change to the file system, which could be as simple as the live OS auto-mounting the device partitions and updating the journal, will change the file system hash.

In addition to standard reporting requirements, as defined by the practitioner's organization, Android forensic investigations should include information on the changes that were made to the original evidence item (mobile device) and any changes that were made should be fully accounted for. This may mean that a practitioner provides details of the technique that was used to collect, examine and analyze the evidence. Practitioners may also include data that they derived from app analysis to remove any ambiguity in the terms they are reporting from app data files and databases.

## 4 CONCLUSION

In this book chapter, we proposed a methodology for collection of evidential data from Android devices. To ensure forensic soundness, the methodology was designed to make as few changes to the evidence source device as possible and when changes were made, they were discrete and, as such, could be easily accounted for by the investigating forensic practitioner. After device identification and preservation techniques (such as ensuring the device was radio suppressed to prevent remote wiping) were undertaken, the first technique in our methodology is to setup the device so that we can boot a live collection OS from the devices volatile memory (RAM).

The live OS is used for collection purposes to ensure that a copy of the device's partitions can be obtained without modifying the contents of the flash memory. Often to collect an image using the device's stock OS requires the device to be rooted, which generally has unknown consequences from a forensic soundness point of view. The live OS may also be used separately, if necessary, to modify the system partition and gain access to any protected data (such as accounts information).

In this sense, the Android live OS is similar to live OS's used on PCs for forensic triage. We recommend that practitioners design their own live OS where possible, to ensure that its operation on the device can be entirely accounted for, or use a live OS from a trusted source (such as from the law enforcement community) where this is not feasible. The practitioner should consider an appropriate method for transport of the physical forensic image collected from the mobile device to an investigation PC when designing their live collection OS. We propose the use of an ADB forwarding function that appears (based on cryptographic hashes matching) to result in a bit-for-bit copy of the device flash partitions being transferred between the mobile device and a PC attached via USB.

Once a physical image of the relevant device partitions has been obtained, a practitioner can commence analysis of the data stored, particularly within the "userdata" partition. Practitioners should commence by examining the files stored by the app in their private storage directories. Once these files have been examined and their type, format and purpose is known, the practitioner should commence examining the files stored, by apps of interest, on external storage. If databases were located as part of file examination, then the next step would be for the practitioner to analyze the databases. Databases often contain metadata that can be of particular interest in forensic investigations. Once all of the apps' files and databases have been explicated, the practitioner can undertake techniques (if necessary) to extract protected AccountManager data from the device. This is achieved by modifying the "system" partition to grant APK packages, signed by a key held by the practitioner, the ability to impersonate other apps on the device to access their secure accounts storage.

Finally, after the practitioner has collected, examined and analyzed all of the data on the device, the majority of evidential data should be available for reporting and presentation. However, if any doubts remain as to the provenience or meaning of the data examined, the practitioner may need to undertake further app analysis. This involves analyzing the memory heaps and APKs for the relevant apps of interest to understand how they operate and ultimately under what circumstances the data located by a practitioner is created. For example, a practitioner may wish to determine how timestamps are updated in the apps databases or how the app authenticates with its hosting service.

We demonstrate the utility of this methodology using seven popular cloud apps in the next chapter [4].